\documentclass[10pt,twocolumn]{article}
\usepackage{simpleConference}
\usepackage{times}
\usepackage{graphicx}
\usepackage{amssymb}
\usepackage{url}
\usepackage{colortbl}
\usepackage{booktabs} 
\usepackage{multirow}
\usepackage{tabulary}
\usepackage[table,xcdraw]{xcolor}
\usepackage{enumitem}

\usepackage{booktabs} % For formal tables
\usepackage[colorinlistoftodos]{todonotes}
\usepackage{graphicx}

\begin{document}

\title{Beyond A/B Testing: Sequential Randomization for Developing Interventions in Scaled Digital Learning Environments}

%\titlenote{Produces the permission block, and
%  copyright information}

 \author{Timothy NeCamp\\
Department of Statistics 
 \\ University of Michigan 
 \\ Ann Arbor, Michigan, USA
\\tnecamp@umich.edu
\\
\\
Josh Gardner\\
Paul G. Allen School of Computer Science \& Engineering \\ University of Washington
 \\ Seattle, WA, USA
 \\
 \\
 Christopher Brooks\\
School of Information
 \\ University of Michigan 
 \\ Ann Arbor, Michigan, USA}
 
 \maketitle

\begin{abstract}
Randomized experiments ensure robust causal inference that is critical to effective learning analytics research and practice. However, traditional randomized experiments, like A/B tests, are limiting in large scale digital learning environments.
While traditional experiments can accurately compare two treatment options, they are less able to inform how to adapt interventions to continually meet learners' diverse needs. In this work, we introduce a trial design for developing adaptive interventions in scaled digital learning environments -- the sequential randomized trial (SRT). With the goal of improving learner experience and developing interventions that benefit all learners at all times, SRTs inform how to sequence, time, and personalize interventions. In this paper, we provide an overview of SRTs, and we illustrate the advantages they hold compared to traditional experiments. We describe a novel SRT run in a large scale data science MOOC. The trial results contextualize how learner engagement can be addressed through culturally-targeted reminder emails. We also provide practical advice for researchers who aim to run their own SRTs to develop adaptive interventions in scaled digital learning environments.
%\footnote{This is an abstract footnote}
\end{abstract}

\section{Introduction}\label{sec:introduction}

In order to continually improve learner experience and maintain engagement, scaled digital learning environments (SDLEs), such as Massive Open Online Courses (MOOCs), intelligent tutoring systems, and open-ended digital educational games, utilize \textit{interventions}. Interventions are modifications made to the learning environment or learners' experience of it, including changing current course content, prompting users to return to the course, or providing additional learning resources (e.g.,\cite{Davis_2018}). It is common to find interventions which are technological, pedagogical, or programmatic in their implementation.

In SDLEs, there is typically diversity both \textit{between} learners (e.g., different learners have different needs) \cite{Kizilcec_2017}, and \textit{within} a learner over time (e.g., a learner's engagement may change throughout a course) \cite{Kizilcec_2013}. To accommodate this diversity, interventions should be adaptive. \textit{Adaptive interventions} can change based on the type of learner (to account for diversity between learners) and change as a learner progresses through the course (to account for diversity within a learner) \cite{almirall_2018}. For example, consider an intervention which introduces review problems throughout a course. The frequency of review problems might adapt based on timing (e.g., learners may not need to review as often in later weeks). The frequency of review might also adapt based on learner performance, building upon work on theories of spaced repetition \cite{reynolds1964effects}.

Due to the large number of users and ease of content manipulation in SDLEs, randomized controlled trials, such as A/B tests, are often used to evaluate intervention options. Typically, these experiments are single randomized trials where each subject is randomized once, and assigned to a single intervention option for the entire trial. However, when the goal is to design a high-quality adaptive intervention in SDLEs, researchers may have important questions about the sequencing, timing, and personalization of intervention options which cannot be answered by A/B tests.  

In this work, we discuss and demonstrate the advantages of an experimental design for developing high-quality adaptive interventions in SDLEs: the \textit{sequential randomized trial} (SRT). In SRTs, a subject is randomized several times to different intervention options throughout the experiment. Sequential randomization is beneficial over one-time randomization for several reasons.  Firstly, by re-randomizing, subjects receive a variety of intervention sequences, and these various sequences can be compared to discover the optimal intervention sequence.  Secondly, instead of only being able to assess the overall effect of receiving one particular treatment, sequential randomization lets researchers discover effects at smaller time-scales (e.g., treatment A does better in week 2 of the course, but treatment B does better in week 3 of the course). These discoveries inform at what time points certain interventions are most effective. Thirdly, re-randomization permits the discovery of important variables measured throughout the course for adapting and personalizing an intervention. As opposed to only being able to discover baseline personalization variables (e.g., treatment A works better for women), we can also discover mid-course personalization variables (treatment A works better for subjects who have been active in the course during the previous day).

The paper is structured as follows: We provide an overview of related prior work on experimentation and personalized interventions in SDLEs in Section \ref{sec:prior-work}. In Section \ref{sec:seqential-randomization}, we formally introduce SRTs and compare them to other common trial designs.
 In Section \ref{sec:srt_examples}, to motivate the design we describe a novel
 %\footnote{The authors are not aware of any previous SRT designs being used in MOOC research to date.}
 SRT performed in MOOCs, the Problem-based Email Reminder Coursera Study (PERCS). 
 This was an experiment aiming to
 improve student retention in an online course by developing a high-quality adaptive email intervention which 
 leverages aspects of cultural diversity and inclusion \cite{Aceves_2014a}. This case study both serves to illustrate the advantages of SRTs and provide context regarding implementation and analysis. Section \ref{seq_time_pers} details three specific advantages of running SRTs. These advantages are exemplified by showing specific results from PERCS. We conclude by providing some practical recommendations for researchers designing their own SRTs in Section \ref{sec:implications-for-practice}.

\section{Prior Work}\label{sec:prior-work}

\subsection{Adaptive Interventions}

Adaptive interventions have been shown to be useful in SDLEs, and have been used extensively in web-based adaptive educational systems \cite{Brusilovsky_2003}. For instance, in \cite{Pardos_2017}, learners were provided personalized content recommendations based on their clickstream data, and in \cite{Davis_2018b}, learners were provided personalized feedback on their learning plans. Adaptive sequences of interventions have also been developed in MOOCs. For example, in \cite{David_2016}, sequences of problems were adapted based on predictions of student knowledge acquisition. Similarly \cite{Davis_2018} chose quiz questions based on course content accessed previously by the learner. While these are only a few examples of adaptive interventions in large scale learning environments, they motivate our desire to improve the process by which such interventions are developed.

In the work discussed, adaptive interventions were developed using learning theory and intention \cite{Davis_2018}, or prediction and machine learning \cite{Pardos_2017, Yu_2017, David_2016}.  In all examples, experimentation was not used to develop the adaptive intervention. In some cases \cite{Davis_2018, David_2016}, experimentation was used for evaluating the adaptive intervention. In these cases, an A/B test is used to compare the designed intervention to a control.

Reinforcement learning techniques, such as multi-armed bandits and contextual bandits, are a type of adaptive intervention which combine exploration (often done through implicit experimentation) and optimization. They use real-time data to learn how to adapt and have also been shown to be useful in SDLEs \cite{Liu_2014b, Clement_2013, Segal_2018, Rafferty_2018}.

\subsection{Experimental Designs}

Experimentation in SDLEs is a common tool for evaluating interventions. 
Unlike quasi- or natural experimental settings \cite{gonzalez_2015, Mullaney_2015}, by randomly assigning interventions, effects of interventions can be separated from effects of confounders (variables that relate both to the type of treatment received and to subsequent outcomes).

A/B tests are a valuable experimental design for improving content and evaluating interventions in MOOCs \cite{savi_2017}. In \cite{Renz_2016}, for example, A/B tests evaluated emails and course on-boarding to improve learner engagement and prevent dropout. In \cite{Davis_2016}, A/B tests were used to test the effectiveness of self-regulated learning strategies. Kizilcec and Brooks \cite{Kizilcec_2017} survey prior work utilizing A/B tests in MOOCs to evaluate nudge interventions and test theory-driven course content changes.

There has also been considerable work in other types of experimental designs beyond A/B testing in SDLEs. \textit{Factorial designs}  \cite{Lomas_2013}  are common ways to evaluate multiple experimental factors simultaneously. 
%Factorial designs are still single randomized trials, however, factorial designs can be incorporated into SRTs, as we demonstrate in the design of PERCS in Section \ref{sec:percs}. 
\textit{Automatic experimentation}  \cite{Liu_2014}, where algorithms are used to search through different intervention options, is another alternative to A/B testing. Though automatic exploration of intervention options may be more efficient, intervention options are still evaluated with single randomized trials. \textit{Adaptive experimental designs} change the randomization probabilities to favor efficacious treatment, with the goal of both evaluating treatment and helping learners currently in the trial \cite{Chow_2008}. Adaptive designs have been used in SDLEs \cite{Williams_2018}. We address the differences between SRTs and these other kinds of designs in Section \ref{sec:compare}.

\section{Sequential Randomized Trials} \label{sec:seqential-randomization}
\subsection{An Overview} \label{sec:overview_srt}

SRTs are trials where an individual is randomized multiple times throughout the course of the trial. Suppose there are two intervention options, such as learners receiving videos taught by a female instructor (intervention A) or a male instructor (intervention B). The simplest example of a SRT would be: During week 1, users have a 50\% chance of receiving intervention A and a 50\% chance of receiving intervention B.  During week 2, users are re-randomized to another treatment. They again have a 50\% chance of receiving either intervention A or B, independent of their week 1 treatment or activity. Hence, about 25\% of users will have received one of each sequence (A, A), (A,B), (B,A), or (B,B), where the parenthetical notation means (week 1 treatment, week 2 treatment).

This simple SRT can be modified for both practical and scientific reasons. Common modifications include using different time durations (e.g., re-randomize every month), increasing the number of time points (e.g., each person is randomized 10 times instead of twice), changing the number of treatments (e.g., A vs B vs C instead of A vs B, or A vs B in week 1 and C vs D in week 2), and altering the randomization scheme (e.g., not having uniform randomization probabilities each week).

SRTs \cite{Lavori_2004, Lavori_2000} have become increasingly common in clinical settings \cite{Lei_2011} but are less common in educational settings \cite{almirall_2018, Kasari_2014} and are even rarer in SDLEs.  Two of the most common types of SRTs are Sequential Multiple Assignment Randomized Trials (SMARTs) \cite{Murphy_2005}, and Micro-randomized Trials (MRTs) \cite{Klasnja_2015}. 

SMARTs are often used in settings where, for either practical or ethical reasons, future randomization and treatment assignment should depend on response to prior treatment.
For example, in a prototypical SMART, individuals are first randomized to one of two different treatment options. Then, at a pre-specified decision point following initial intervention, users who did not respond to initial treatment are re-randomized, while users who did respond to initial treatment are not re-randomized and continue with the efficacious initial treatment \cite{Pelham_2016}. 

MRTs are useful for interventions that can be delivered quickly and frequently (such as delivering text messages or notifications to a subject's phone). Typically, the goal of an MRT is to estimate the short-term effect of these interventions and understand how that effect depends on time and context. MRTs have been mostly used in the mobile health space \cite{Klasnja_2015}. Due to the high frequency of intervention delivery, users in an MRT are typically re-randomized hundreds or thousands of times.

\subsection{Comparisons to Other Designs}\label{sec:compare}

\subsubsection{Single Randomized Trials}

Single randomized trials, such as A/B tests and factorial designs, are trials where subjects are randomized \textbf{one time}. In A/B tests (often called a 2-arm randomized controlled trial in healthcare and education), each subject is randomized one time to either intervention A or intervention B. Factorial designs are an extension of A/B tests where each subject is randomized one time to several intervention components simultaneously. SRTs differ from single randomized trials because in SRTs, subjects are randomized several times, causing them to receive different treatments at different times throughout the trial. 

Single randomized trials can still be used to evaluate adaptive interventions. For example, if treatment A is defined as an adaptive intervention (e.g., a fixed sequence of different intervention options) and treatment B is defined as a control, an A/B test can compare the adaptive intervention to the control with a single randomization. However, this A/B test is limited in answering questions about sequencing, timing, and personalizing.

A/B tests are often used in confirmatory trials. Confirmatory trials are trials used to ensure strong evidence (or additional evidence) of a treatment's efficacy. In contrast, SRTs are useful as exploratory trials which explore a large number of possible treatment sequences and learn how to adapt those sequences. After running a SRT and developing a high quality adaptive intervention, the intervention can be confirmed in a simple A/B confirmatory trial \cite{Collins_2007}.

SRTs can also be thought of as factorial designs, where users are sequentially randomized to each factor over time. In the simplified SRT example in Section~\ref{sec:overview_srt}, the design can be considered a $2 \times 2$ factorial design where factor 1 is week 1 treatment, and factor 2 is week 2 treatment \cite{Almirall_2014}.

\subsubsection{Online Optimization Designs}

There are other designs (both experimental and not) which aim to optimize intervention delivery while simultaneously collecting data. In adaptive trial designs \cite{Chow_2008}, randomization probabilities are changed throughout the trial in order to both provide efficacious treatment to users, and still obtain good estimates of treatment effects. 
Online reinforcement learning methods (such as multi-armed bandit and contextual bandit algorithms) can also be used for optimizing intervention delivery \cite{Liu_2014b, Clement_2013, Segal_2018, Rafferty_2018}. 

SRTs are distinctive from adaptive trial designs and online reinforcement learning methods since they do not use data collected during the trial to change randomization schemes. In reinforcement learning language, SRTs can be seen as pure exploration with no exploitation. There are advantages of not using earlier trial data to inform future treatment allocation:
\begin{enumerate}
\item Using online optimization techniques can cause bias in estimating treatment effects \cite{Nie_2017, Rafferty_2018}. These bias issues do not arise in SRTs.

\item SRTs provide rich exploratory data for discovering which variables are valuable for informing treatment decisions, making them useful when these variables are unknown (see Section \ref{sec:personalization-with-srts}). In contrast, many reinforcement learning algorithms, such as contextual bandits \cite{Lan_2016}, require these variables to be pre-specified.

\item SRTs can actually utilize reinforcement learning methods. Batch off-policy reinforcement learning algorithms (such as Q-learning) can be applied to SRT data to discover an optimal adaptive intervention, as in \cite{Zhao_2009}.
     
\end{enumerate}

\section{Applications of Sequential Randomized Trials} \label{sec:srt_examples}

SRTs can inform a large variety of interventions including course content sequencing, course material manipulations, and learner nudges (such as encouraging messages and problem feedback). We highlight three examples. The first two examples are hypothetical scenarios to demonstrate different types of possible SRTs. The third example, PERCS, is a SRT run in a data science MOOC and is the main working example. 

\subsection{Video Optimization} For each video in a MOOC, there are two versions, one video which shows only slides and one which shows the instructor's head occasionally interspersed with the slides \cite{Guo_2014}. Researchers are unsure about which sequence of videos are better. 
Learners might prefer only slides, some might prefer those with instructors, or some might prefer a variety of videos. Also, there may be learner characteristics that affect learner preference. For example, learners who initially performed poorly after watching an instructor video might be better off seeing a slide-only video. To provide insight into these questions and hypotheses, a researcher could run a SRT: When a learner enters a course, they are randomized initially to receive an instructor video or a slide only video. Then, after watching the first video, they are re-randomized to either instructor or slide video for the next video they watch. This continues through the entire course.

\subsection{Content Spacing} Researchers are often unsure about optimal review problem sequencing to maximize knowledge retention while minimizing review time \cite{reynolds1964effects}.  Learners may benefit from frequent review at the beginning, with less frequent review later. These benefits may also be dependent on certain learner characteristics. For example, poor-performing learners may benefit from more frequent review.
A SRT can answer these questions: For a problem recommendation system, a learner starts the recommender for a time window (e.g., 50 problems). Then, after this time period, every learner is randomized to one of 3 groups: no review, minimal review, large review. The grouping determines how often they receive previously-seen problems (for review) during the next 20 problems. If a learner is in the no review, minimal review, or large review group they will receive previously-seen problems 0\%, 5\%, or 20\% of the time, respectively. After completing the next 20 problems, each user will be re-randomized to one of the 3 groups. This randomization scheme continues. After every 20 problems completed, a user is re-randomized to one of the 3 groups.

\subsection{Problem-based Email Reminder Coursera Study (PERCS)}\label{sec:percs}

\subsubsection{Motivation}
\label{sec:rqs}

A well-known challenge in MOOCs are the low completion rates. While there are many factors contributing to MOOC dropout \cite{Greene_2015}, the goal of PERCS is to determine whether dropout can be ameliorated by using weekly email reminders to motivate learners to engage with course content. Our context of inquiry was the Applied Data Science with Python Coursera MOOC, taught by Christopher Brooks. Weekly emails were sent to learners and may have contained one or more of several factors intending to impact learner engagement (for an example, see Figure~\ref{fig:example-email}):
\begin{enumerate}
    \item The email could have contained a motivating data science problem to challenge the user to learn the upcoming week's content. This factor was based on evidence from the problem-based learning literature suggesting that situating instruction in the context of problems is an effective way to engage learners \cite{barrows1985design}.
    \item  The email might also have contained a location-specific primer and a data science problem relevant to that user's specific culture (e.g., an Indian user might receive a problem about Bollywood or weather patterns in India). This factor was based on the work in the culturally relevant and culturally responsive pedagogy communities, where situating instruction in a manner that considers the local context is seen as beneficial \cite{Aceves_2014a}.
    \item The email may have utilized growth mindset framing \cite{Dweck_2008}, a psychological framing method used to support learning. While growth mindset has been heavily studied, its effects are in dispute \cite{Kaijanaho_2018}, and growth mindset framing has seen only limited application in SDLEs.
\end{enumerate}

Given these weekly email options, we hope to develop an adaptive intervention of weekly email reminders to increase engagement and reduce dropout. A high-quality adaptive intervention should sequence emails in a way that promotes engagement continually through the course.  Since different emails may be more or less effective during different weeks in the course, each week, the intervention should send the most effective email for that week. Finally, since we might not expect the same email to work well for everyone, the intervention should also adapt to the learner's current course behavior. In order to develop a high quality adaptive intervention of emails, we need to answer the following research questions:

\begin{enumerate}[label=RQ\arabic*]
    \item \textbf{Sequencing}: Which sequence of emails most improves course activity in later weeks?\label{rq1}
    \item \textbf{Timing}: Which email problem type is most effective, on average, for bringing learners back to the course during each week? \label{rq2}
    \item \textbf{Personalization}: Are certain data science problem emails more or less effective for active learners? \label{rq3}
\end{enumerate}
            
\subsubsection{Design}

\begin{figure}
    \centering
    \includegraphics[width= 0.7 \columnwidth]{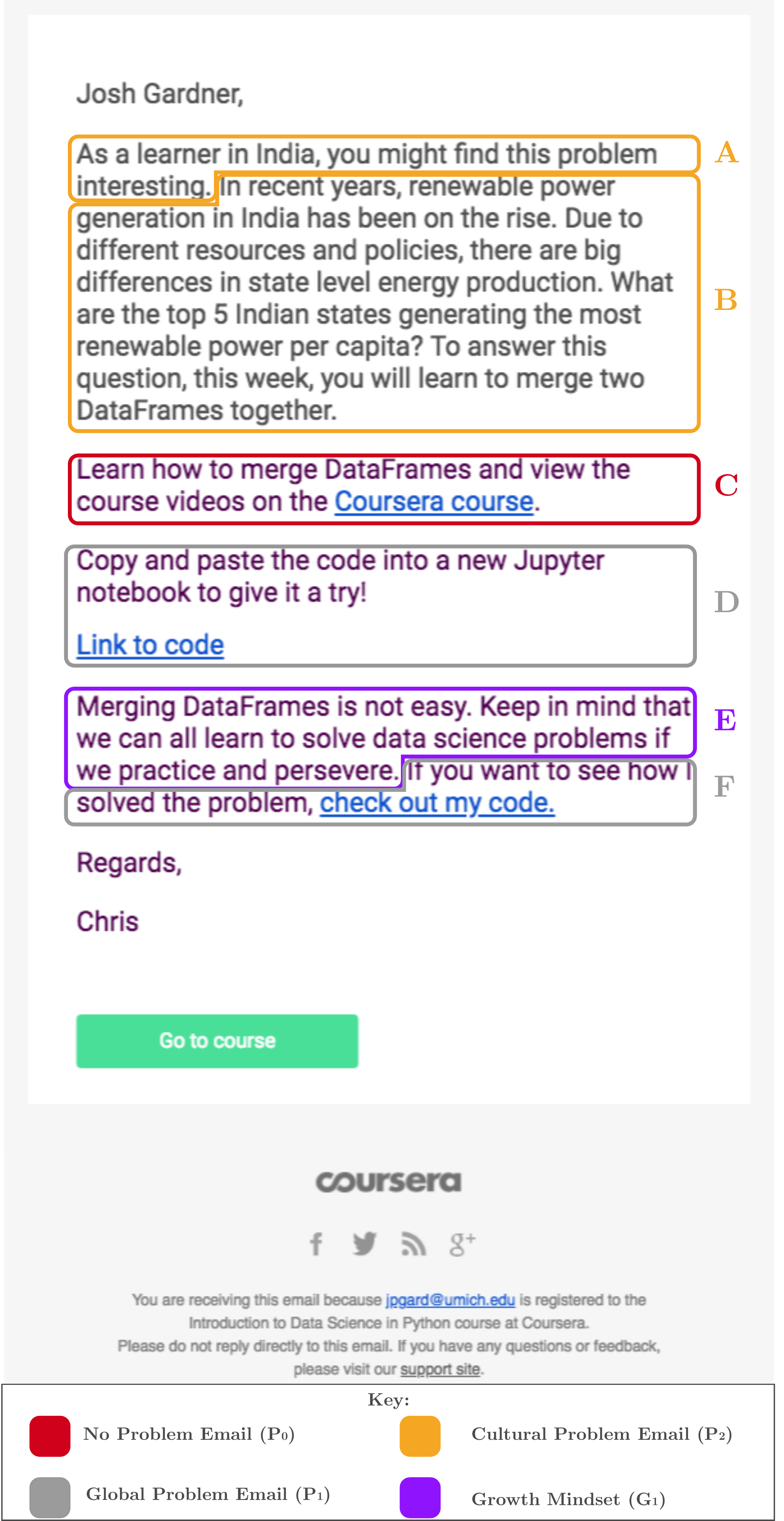}
    \caption{Example email structure using content which is populated based on their assigned treatments. (A) Identity activation prompt. (B) Culturally-relevant problem using data related to geographic identity. (C) Reminder to return to course. (D) Link to problem code. (E) Growth mindset framing. (F) Link to problem solution. Elements of the same email type are grouped by color (see key).}
    \label{fig:example-email}
\end{figure}

\begin{figure*}[h!]
    \centering
    \includegraphics[width = 0.7 \textwidth]{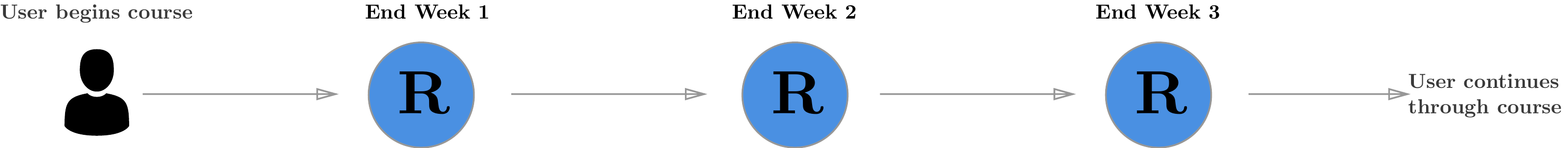}
    \caption{PERCS trial design. ``R'' indicates stages where randomization is conducted according to the probabilities in Table \ref{tab:probs}.}
    \label{fig:percs_design}
\end{figure*}

\begin{table*}[h!]
\begin{tabular}{ll|
>{\columncolor[HTML]{C0C0C0}}l |l|l|l|}
\cline{1-1} \cline{3-6}
\multicolumn{1}{|l|}{\cellcolor[HTML]{C0C0C0}{\color[HTML]{000000} \textbf{No Email $E_0$}}} &  & \textbf{Email $E_1$}                                                          & \cellcolor[HTML]{C0C0C0}\textbf{\begin{tabular}[c]{@{}l@{}}No Problem \\ Email ($P_0$)\end{tabular}} & \cellcolor[HTML]{C0C0C0}\textbf{\begin{tabular}[c]{@{}l@{}}Global Problem \\ Email ($P_1$)\end{tabular}} & \cellcolor[HTML]{C0C0C0}\textbf{\begin{tabular}[c]{@{}l@{}}Cultural Problem \\ Email ($P_2$)\end{tabular}} \\ \cline{1-1} \cline{3-6} 
\multicolumn{1}{|l|}{0.14 ($T_1$)}                                                           &  & \textbf{\begin{tabular}[c]{@{}l@{}}No-Growth \\ Mindset ($G_0$)\end{tabular}} & 0.14 ($T_2$)                                                                                         & 0.14 ($T_3$)                                                                                             & 0.14 ($T_4$)                                                                                               \\ \cline{1-1} \cline{3-6} 
                                                                                             &  & \textbf{\begin{tabular}[c]{@{}l@{}}Growth \\ Mindset ($G_1$)\end{tabular}}    & 0.14 ($T_5$)                                                                                         & 0.14 ($T_6$)                                                                                             & 0.14 ($T_7$)                                                                                               \\ \cline{3-6} 
\end{tabular}
\caption{An overview of the probability learners will be assigned to a treatment $T_n$. Individual treatments are shown in white cells, while groups of treatments are referred to by the tab row or column headers (e.g., all cultural problem emails as $P_2$)}
\label{tab:probs}
\end{table*}

A sequentially randomized factorial trial design was an effective method to jointly address the main research questions of PERCS. At the end of weeks one, two, and three of the four-week long MOOC, learners were randomly assigned to receive one of four different email categories: an email message with a problem that reflects their geo-cultural situation based on IP address (\emph{cultural problem email}), an email with a generic non-culture specific problem (\emph{global problem email}), an email with \emph{no problem}, or \emph{no email} at all. Those who received an email were uniformly randomly assigned to have the email be framed with or without growth mindset. 

The growth mindset factor crossed with the three email categories makes each week of PERCS a $2 \times 3$ factorial design with an additional control condition of no email. See Table~\ref{tab:probs} for each week's factorial design and randomization probabilities.

Figure \ref{fig:example-email} illustrates an example email. The emails were developed in a ``cut and paste'' format: When adding in a condition to an email (e.g., growth mindset framing or adding a problem), we do not change other aspects of the email, and simply insert text from the relevant condition into a consistent email template. Using the cut and paste across different conditions allows us to attribute treatment effects purely to the condition being added, and not other aspects of the email. Emails are delivered directly using the Coursera platform's email capabilities for instructors.
 
The most novel aspect of PERCS is the sequential randomization. That is, a particular learner is not assigned to a single fixed email condition for all three weeks. Instead, as shown in Figure \ref{fig:percs_design}, a learner is re-assigned (with the same randomization probabilities) to a different email condition each week. The randomizations across weeks are independent, hence in PERCS, there are $7^3$ different possible email sequences students may receive.

\subsubsection{Notation}

Throughout the rest of the paper, we will be referring to specific treatments, groups of treatments, and sequences of treatments in PERCS. We introduce notation to ease our description. As shown in Table \ref{tab:probs}, $T_1, T_2, \dots, T_7$ refers to a particular email. For example, $T_5$ is an email containing no problem, and growth mindset framing. To refer to groups of emails, we used labeling contained in the column and row headers in Table \ref{tab:probs}. $E$ is used to group together conditions where any email was sent ($E_1$) vs no email being sent ($E_0$). So users in $E_1$ refer to any user receiving emails $T_2$ through $T_7$. $G$ is used to group together growth mindset emails ($G_1$) and non-growth mindset emails ($G_0$). Users in $G_0$ refer to any user receiving emails $T_2, T_3, $ or $ T_4$. $P$ is used to group together no problem emails ($P_0$), global problem emails ($P_1$), and cultural problem emails ($P_2$). Thus users in $P_1$ are any users receiving emails $T_3$ or $T_6$.  Lastly, we use parenthetical notation to describe the sequences of emails over the three weeks, i.e., (week 1 treatment, week 2 treatment, week 3 treatment). As an example, we would refer to users who received a global problem email in week 1, any email in week 2, and a no-growth mindset email in week 3 using ($P_1, E_1, G_0$).

\subsubsection{Experimental Population}

We focused our trial on the two largest populations of learners enrolled in the course as determined by IP address, Indian and US-based learners. All Indian and US learners who signed up for the Applied Data Science with Python Coursera MOOC between April 1 and June 10, 2018 participated in PERCS. A total of 8,681 unique learners (3,455 Indian, 5,226 US) were sent 22,073 emails. 
%% india 7717 3455
%% us 12,356 5,226

\subsubsection{Single randomized version of PERCS: PERCS-AB}

To highlight the advantages of sequential randomization, PERCS will be compared to the single randomized version of PERCS, PERCS-AB. PERCS-AB has the exact same randomization probabilities as PERCS, however in PERCS-AB, learners are randomized only at the end of week 1 to one of the 7 email types. They are then sent that exact same email type in weeks 2 and 3. Hence, there are only 7 possible sequences ($T_1, T_1, T_1)$, $(T_2,T_2,T_2)$, $\dots$,  $(T_7,T_7,T_7)$.

\section{Sequencing, Timing, and Personalizing Interventions through SRTs} \label{seq_time_pers}.
%or developing high quality adaptive interventiosn wtih srts

Next we highlight how SRTs can provide answers to questions regarding sequencing, timing, and personalizing interventions. For each type of question, we provide (1) an overview of the question and contextualize it within PERCS, (2) evidence of why SRTs are beneficial for answering that question as illustrated through comparing PERCS and PERCS-AB, (3) further discussion, and (4) answers to the question in the context of PERCS.

\subsection{Sequencing} \label{sec:optimal-sequence}

\subsubsection{Overview} 
When many different intervention options can be delivered at different times, proper sequencing of interventions is critical. An intervention that worked at one time may not work at a later time. Also, receiving the same intervention multiple times may be less effective than receiving a variety of interventions. Researchers may not know which sequence of intervention options will lead to the best outcomes for learners. For PERCS, \ref{rq1} refers to sequencing-- are there sequences of emails which improve course activity in the later weeks?

\subsubsection{Advantages of SRTs}

SRTs provide data that allows experimenters to compare different sequences of interventions. By re-randomizing learners, learners receive a variety of treatment sequences. In PERCS, for example, with the large sample size, there are learners randomly assigned to each of the $7^3$ possible sequences of treatments (e.g., $(T_1, T_1, T_1)$, $(T_1, T_1, T_2)$, $(T_1, T_2, T_2)$, or $(T_2, T_1, T_2)$).

By randomizing learners to each possible treatment sequence, researchers are now able to compare these treatment sequences. For example, researchers might hypothesize that learners only need a cultural problem email in the first week to believe the content is inclusive. They could then compare the course activity of users initially receiving a cultural problem email followed by two global problem emails $(P_2, P_1, P_1)$ vs users receiving a cultural problem email all three weeks $(P_2, P_2, P_2)$. When thinking of PERCS as a $7 \times 7 \times 7$ factorial design (where each week's treatment is a different factor), comparing sequences is analogous to simple effects analysis in factorial designs.

In A/B tests, since learners are not re-randomized, only sequences where each learner received the same treatment every time can be compared. In PERCS-AB, learners are only randomized to 7 possible sequences and thus comparisons can only be done between these 7 sequences. If there is any benefit to receiving different interventions (and/or in a different order) then this could not be discovered by PERCS-AB. However one would note that all of the comparisons of PERCS-AB can be done with data collected through PERCS, coming at the cost of a reduced sample size. This tradeoff demonstrates again why SRTs are especially well suited for MOOC experimentation, where there is a large number of diverse participants (and thus a broad exploration might be suitable).
 
\subsubsection{Discussion}
%Comparing sequences of treatment groups

Often times, researchers are not interested in such specific sequence comparisons. Instead, they are interested in questions about sequences of groups of interventions. In this case, one could perform similar comparisons, but combining users over all of these groups. For example, in PERCS,
we can assess how often reminder emails should be sent. Is it better to space emails out weekly or bi-weekly?

To answer this question, we could compare course activity of users receiving the sequence which spaces emails out by 2-weeks $(E_1, E_0, E_1)$, to the sequence which sends emails every week $(E_1, E_1, E_1)$. The sample size for this comparison will be much larger than for the individual sequence comparisons.

In addition to comparing groups of sequences, researchers might only be interested in comparing sequences for a small number of time points. For example, we could use data collected in PERCS to understand if the effects of week two problem-based emails on course activity were different based on what type of problem-based email the user had received in week one. Specifically, does receiving no email vs cultural problem-based email in week one change the benefits of receiving a cultural problem email in week two?  To answer this question, we compare course activity in week two of users receiving sequences $(E_0, P_2, $any) vs ($P_2, P_2, $any). Notice we are not concerned with week three assignment, so we include sequences with any treatment in week three (i.e., any includes $T_1, T_2, \dots, T_7$).

\subsubsection{Results from PERCS} \label{results_seq}
\begin{figure}
\includegraphics[width= \columnwidth]{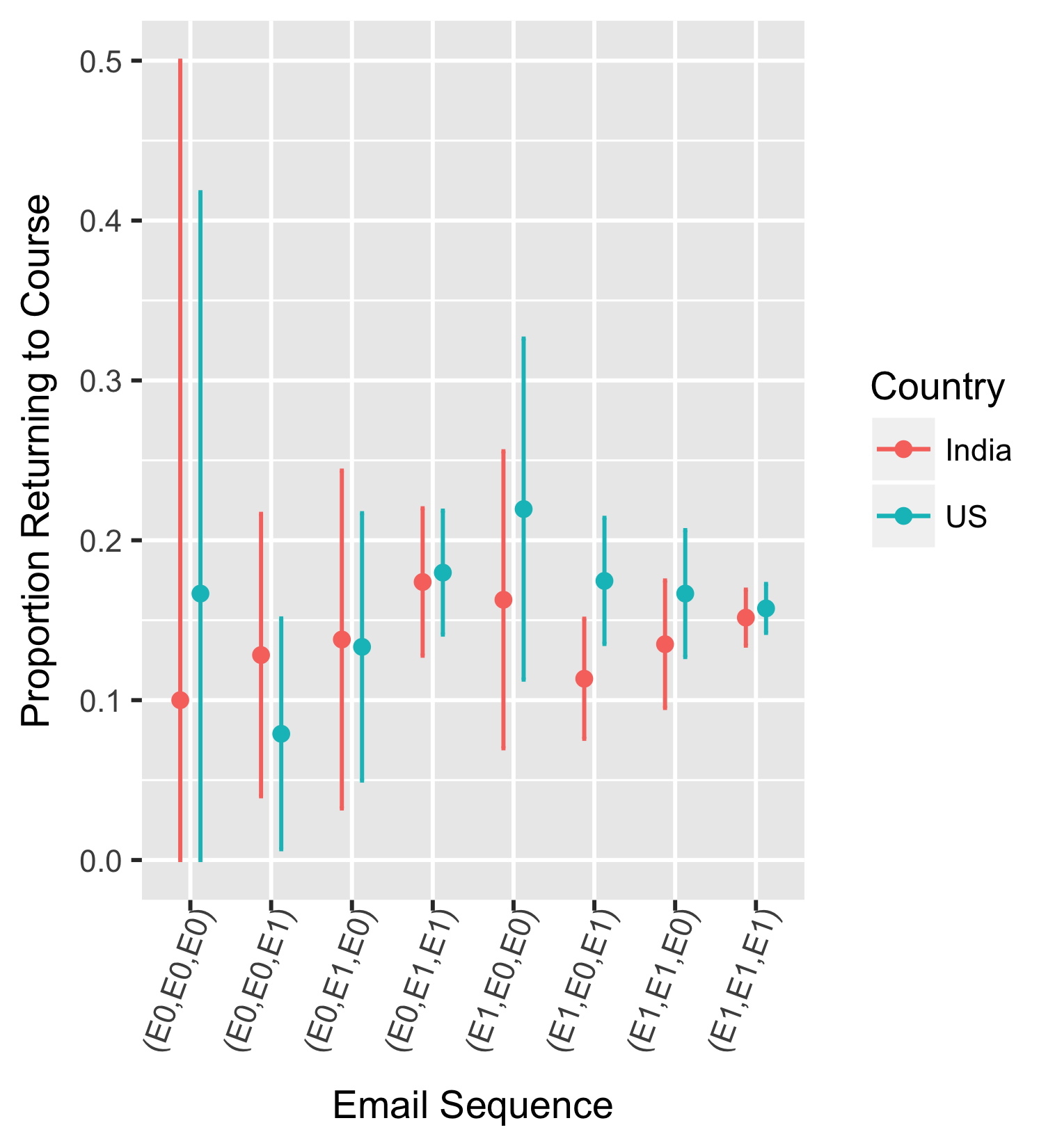}
\caption{95\% confidence intervals for the proportion of students returning to the course in week 4 after receiving a given three week sequence of emails ($E_1$) and no email ($E_0$)} \label{fig:seq_results}
\end{figure}

We now present results addressing \ref{rq1}. Since there are a large number of potential sequences, here we focus on the highest-level comparison: understanding the proper sequence of any email and no email. To do the comparison we look at the proportion of students returning to the course in week 4 (the final week) after receiving a given sequence of email ($E_1$) and no email ($E_0$), as shown in Figure \ref{fig:seq_results}.

We note some important observations. For US learners, sequences with emails sent in the first week (i.e., those sequences which start with $E_1$, the last four columns in Figure \ref{fig:seq_results}), tend to be slightly more beneficial than sequences without. For Indian learners this is not true. 

Overall, for both countries, the confidence intervals across sequences are mostly overlapping, indicating that differences in effects of various email sequences are not significant but only suggestive. The size of the confidence intervals change due to the probability of receiving an email ($E_1$) being larger than the probability of not receiving an email ($E_0$), and are not induced by the treatment itself.

\subsection{Timing}\label{sec:avg-treatment-effects}

\subsubsection{Overview}

Instead of being interested in a sequence of interventions, researchers might want to know about the effect of an intervention option at a particular time point. A treatment that was effective at the beginning of the course may be less effective towards the end of the course. By understanding effects of intervention options at a time point, designers can build adaptive interventions which deliver the optimal treatment at all times. For PERCS, \ref{rq2} regards timing-- which email type is most effective during each week?

\subsubsection{Advantages of SRTs}

Answering research questions about treatment timing is done by estimating average treatment effects, i.e., the average effect of a given treatment option at a given time point. Here, the averaging is over all prior treatment received, allowing the effect to be only for the particular time point of interest. When thinking of PERCS as a $7 \times 7 \times 7$ factorial design (where each week's treatment is a different factor), estimating average treatment effects is analogous to main effects analysis in factorial designs.

SRTs permit the estimation of average treatment effects at various time points. By re-randomizing individuals, we can separately estimate average treatment effects at each time point and eliminate dependence on treatment delivered previously.

To exemplify, in PERCS, to understand which email type is most effective in week 3, we compare the average effect of cultural problem emails in week 3 (any, any, $P_2$) compared to no email in week 3 (any, any, $E_0$). By doing this comparison, we average over all treatments delivered prior to week 3, and isolate the effect of interest to emails only in week 3. 
By re-randomizing, the individuals receiving a cultural problem email ($P_2$) in week 3 and individuals receiving no email ($E_0$) in week 3 both have, on average, the same distribution of treatments delivered prior to week 3. Hence, the comparison at week 3 is under the same prior treatment distribution for both groups in the comparison.

In PERCS-AB, such a comparison is impossible. Since individuals receive the same email all three weeks, the individuals receiving a cultural problem email in week 3 had a different prior treatment distribution compared to those receiving no email in week 3. This makes the comparison at week 3 implicitly dependent on these different prior treatment distributions.

In SRTs, average treatment effects can also be estimated for outcomes measured \textit{after} the next re-randomization (sometimes called delayed effects  \cite{Murphy_2005}), by averaging over future treatment.  For example, in PERCS (but not PERCS-AB), we are able to estimate the average effect of cultural problem emails in week 2 on week 4 course activity by averaging over week 3 treatment.

\subsubsection{Discussion}
%The Difference Between Average Treatment Effects and Sequence Effects

%% I might be able to cut this

Research questions on average treatment effects often seem similar to questions regarding sequence effects. Three different research questions in PERCS elucidate the differences:

    (1) What is the effect of receiving a cultural problem email in week 3, after not receiving any email prior, ($E_0, E_0, P_2$) vs ($E_0, E_0, E_0$)?
    
    (2) What is the average effect of receiving a cultural problem email in week 3, (any, any, $P_2$) vs (any, any, $E_0$)?
    
    (3) What is the effect of receiving a cultural problem email every week until week 3, $(P_2, P_2, P_2)$  vs $(E_0, E_0, E_0)$?

Questions 1 and 3 are questions of comparing sequences of treatments, while question 2 is about average treatment effects. Note that all three questions can be answered by PERCS, but only question 3 can be answered by PERCS-AB.

Also, one advantage of analyzing average treatment effects is that the sample size is typically larger than when comparing individual sequences of treatments, since the comparison does not restrict users based on what they received before or after the given week of interest.

\subsubsection{Results from PERCS}

\begin{figure}
\includegraphics[width= \columnwidth]{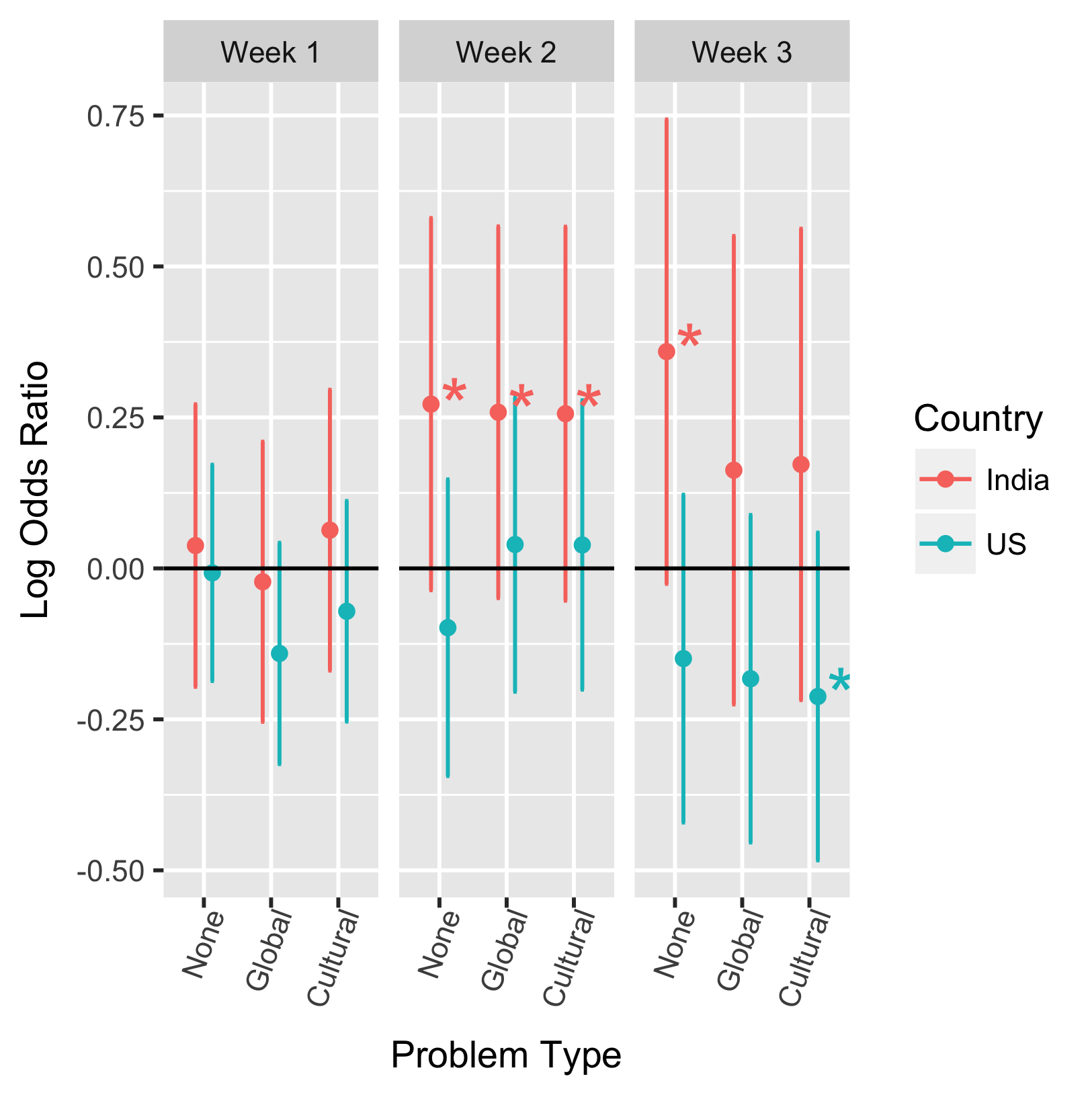}
\caption{95\% confidence intervals of the log odds ratios of probability of returning to the course in the subsequent week for no problem ($P_0$), global problem ($P_1$), or cultural problem ($P_2$) emails, when compared to no email ($E_0$). \newline {\large*} indicates a moderate effect with significance at $\alpha$ = .2}
\label{avg_eff_results}
\end{figure}

To assess \ref{rq2}, for each week, we perform logistic regression with a binary outcome indicating whether the user clicks anything in the course during the week after receiving an email. The results are in Figure~\ref{avg_eff_results}. Negative values indicate a reduced chance of returning to the course, compared to no email ($E_0$), while positive values indicate an increased chance of returning to the course.

The results in Figure~\ref{avg_eff_results} show that the impact of emails on Indian learners in weeks 2 and 3 is largely positive, but the impact of receiving a no problem email ($P_0$) is as good or better than receiving either global or cultural problem emails ($P_1$ and $P_2$, respectively). Also, in week 2, emails of all types were moderately effective, indicating this is a good time to send Indian users email reminders for this course.%, but including motivating data science problems in those emails may not be worth the effort.

For US learners, the effects are non-significant across all emails except for the cultural problem email ($P_2$) in week 3, and the log odds ratios are small and in many cases negative. This indicates that for all weeks, emails for US users did not impact their propensity to return to the course, and may even deter them from returning -- a counter-intuitive, but important, insight regarding timing of communication with learners.

Emails were more effective for Indian learners compared to US learners, despite email open rates being significantly larger for US learners (41.7\% open rate) than Indian learners (27.0\% open rate, p-value < 0.001 for difference in proportions). This suggests that Indian learners may benefit even more if we could increase the email open rate. Also, note that open rates did not differ across email type, as all emails in a given week had the same subject line.

\subsection{Personalizing}\label{sec:personalization-with-srts}

\subsubsection{Overview}
%\TN{This should match beginning}
SDLEs are notable for the high degree of diversity both across and within learners \cite{Kizilcec_2013,Kizilcec_2017}. Due to this diversity, we might expect treatment effects to vary along with relevant learner attributes. If an intervention works for a specific user at a given time, that intervention may not be effective for a different user, or even the same user at a different time.  Personalizing treatment, by discovering when and for whom certain treatments are most effective, is critical. For PERCS, \ref{rq3} regards personalization-- are certain emails more effective for active learners?

In clinical trials, learning how to personalize treatment is synonymous with discovering \emph{moderators} \cite{Kraemer_2002}. Moderators are subject-specific variables which change the efficacy of an intervention. For example, if a MOOC intervention works better for older users than younger users, then age is a moderator and can then be used to personalize; one may only deliver the intervention to older learners. Answering \ref{rq3} in PERCS is equivalent to understanding if previous course activity is a moderator of email effectiveness.
\subsubsection{Advantages of SRTs}

Both SRTs and A/B tests permit the discovery of \emph{baseline moderators} -- variables measured prior to the first treatment randomization (e.g., gender, location, age, scores on early assignments)  which moderate the effect of treatments. Baseline moderators are important for personalizing treatment. However, understanding how to change treatment based on variables measured throughout the course -- \emph{mid-course moderators} -- is critical. Discovering mid-course moderators tells practitioners how to personalize interventions throughout the course to account for heterogeneity both between users and within a user over time. 

Unlike A/B tests, SRTs permit the discovery of mid-course moderators. Statistically, due to bias introduced when including post-randomization variables in analyses, one can only discover moderator variables measured prior to randomization \cite{Kraemer_2002}. If all learners were only randomized once, potential moderators measured after the first treatment cannot be discovered. By re-randomizing in SRTs, moderators measured before each of the randomizations (which now includes mid-course data) can be discovered. 
Because users were randomized again at the end of week three, we can use PERCS data to answer \ref{rq3} about week three emails. Specifically, we can assess how activity during week three moderates the effect of emails sent at the end of week three. In PERCS-AB, week three course activity cannot be assessed as a moderator since it is measured after the one and only randomization in week one. 

\subsubsection{Discussion}

We can evaluate mid-course moderators that include previous treatment. For example, we may think that responsiveness to previous treatment could inform how to personalize future treatment. Those that were responsive should continue receiving the same treatment while those that were non-responsive should receive different treatment. In PERCS, for example, we might expect week two emails to benefit users who responded positively to emails in week one. One could assess this by comparing two groups: Group 1 are users who received an email ($E_1$) but did not click in the course afterwards (i.e., email non-responders). Group 2 are users who received an email ($E_1$) but did click in the course afterwards (i.e., email responders). We could compare the effect of emails in week two for Group 1 vs Group 2.

Also, as learning content optimization starts to happen in real-time (i.e., reinforcement learning), knowing which real-time variables to measure and use for online optimization is critical. SRTs permit this discovery.

\subsubsection{Results from PERCS}

\begin{figure}
\includegraphics[width = \columnwidth]{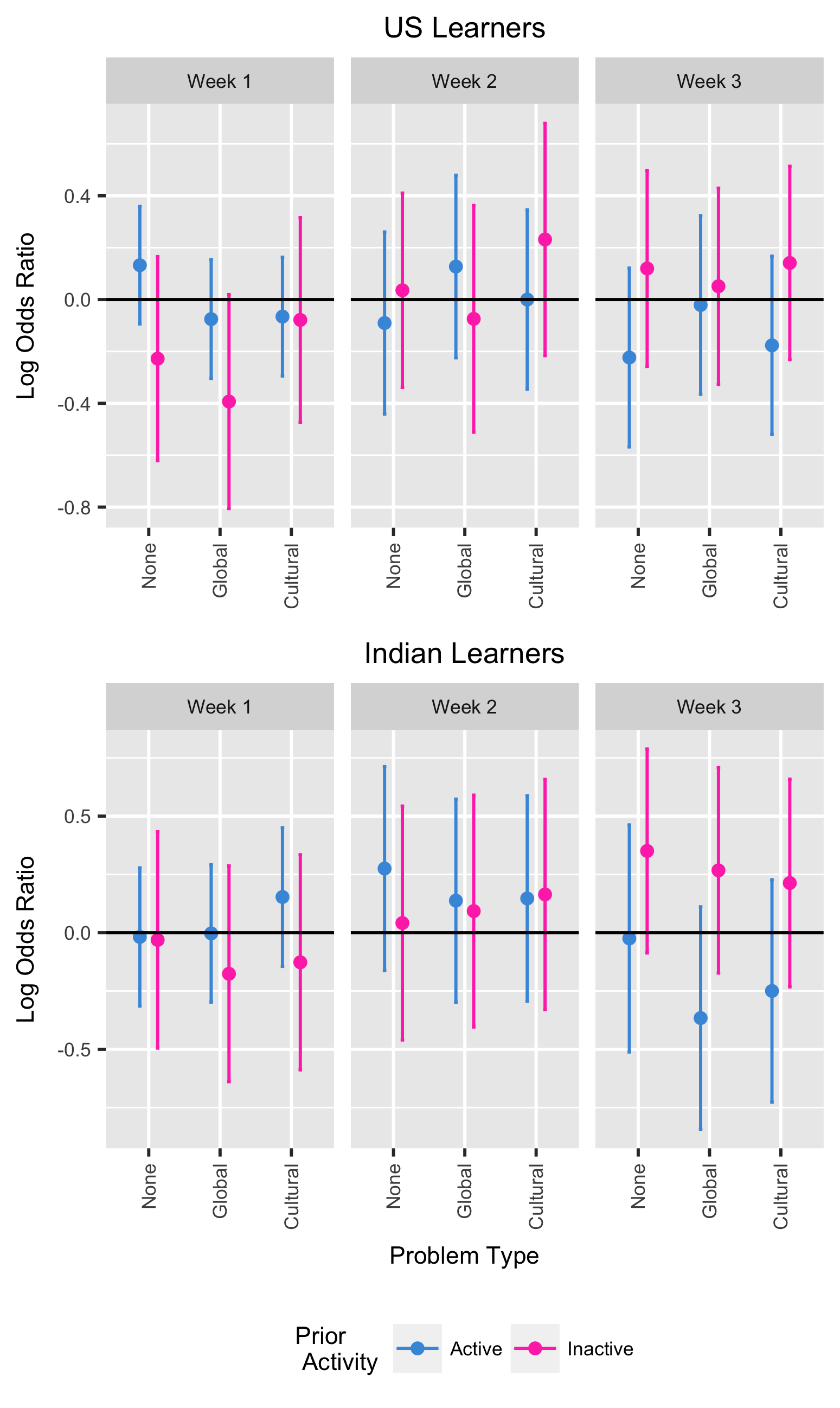}
\caption{95\% confidence intervals of the log odds ratios of probability of returning to the course in the subsequent week for no problem ($P_0$), global problem ($P_1$), or cultural problem ($P_2$) emails, when compared to no email ($E_0$). The log odds ratio is calculated for users who had activity (active) or did not have activity (inactive) in the prior week.} \label{fig:moderator}
\end{figure}

For PERCS, we were most interested in the previous week's course activity as a mid-course moderator. We wanted to see how different types of emails effect active and inactive users differently. Active users are defined as users who had one or more clicks in the course during the previous week. Before the trial, we did not know if problem-based emails would encourage inactive users (because they need the motivational reminder) or discourage inactive users (because the problem may be too advanced). To assess this, in Figure~\ref{fig:moderator} we plot the log odds ratios of the probability of returning to the course in the subsequent week for different email conditions, compared to the control of no email ($E_0$). A positive log odds ratio indicates users had a higher chance of returning to the course after receiving the corresponding email problem type (compared to the no email control).

%There are two key observations drawn from the results in 
In week one, for both US and Indian learners, reminder emails of all types performed better for active users than inactive users (higher log odds for blue than magenta, Figure~\ref{fig:moderator}). In week three, the sign of the moderation switched, as emails performed better for inactive users compared to active users (higher log odds for magenta than blue). This moderation was larger for Indian learners. Also, the log odds ratios for inactive users were positive, while the log odds ratios for active users were negative, suggesting that emails were beneficial for bringing back inactive users, but potentially harmful to active learners.
To ensure we encourage inactive users to return to the course while not discouraging active users, these results suggest that email sending should adapt based on course activity and the course week. Note that the confidence intervals are mostly overlapping, indicating that differences are not significant but only suggestive.

\section{Implications for Practice}\label{sec:implications-for-practice}

This section outlines some useful considerations for researchers who are interested in designing and running SRTs for SDLEs. First, researchers should always remember that experiments are created to inform scientific understanding and answer questions about interventions. Deciding if a SRT is the correct trial design depends purely upon the scientific questions of interest. For example, if researchers are not interested in understanding how to sequence, time, and personalize interventions, running a SRT is unnecessary. In PERCS, if a researcher were interested in only comparing two sequences of interventions then it may be best to run an A/B test on those two sequences and forgo answers to other questions. 

In situations where a SRT is appropriate, there are also important trial design considerations.  All intervention sequences in a SRT should be useful, feasible, and scalable. Intervention sequences which could never be used in practice or which are knowingly deleterious to subjects should be avoided. For example, suppose a researcher was curious about optimal ordering of course content and, in turn, sequentially randomized learners to various sequences of content. If content B requires information taught in content A, none of the sequences in the experiment should place content B before content A. As another example, in PERCS, because the MOOC platform does not currently support fully-automated, scheduled delivery of email, messages were only sent once per week. We did not explore the possibility of sending emails many times per week (e.g., re-randomizing email sending every day) because, in practice, interventions which require manual daily email sending would not currently be feasible for instructors on this platform.

As in any other trial design, sample size and power calculations are important for SRTs. Sample size calculations should be based on the most important research questions the trial intends to answer. Since there are a larger number of possible treatment sequences in a SRT, calculating power and sample size requires researchers to consider which subset of users will be randomized to the sequences of interest. Then, sample size calculations are similar to A/B tests for both comparing interventions and discovering moderators \cite{Oetting_2011}. 
Due to the larger space of potential treatment sequences, sample sizes usually need to be larger for SRTs (compared to A/B tests). To increase power, researchers may consider changing randomization probabilities to favor interventions of interest.

\section{Conclusions, Limitations, and Future Work}

Adapting, sequencing, and personalizing interventions is critical to meet the diverse needs of learners in SDLEs. When designing adaptive interventions, experimentation is useful for evaluating and comparing possible sequences. Unfortunately, single randomized A/B tests cannot answer questions about ways to adapt and sequence interventions throughout a course.  In this work, we demonstrated how a new type of experimental design, SRTs, are valuable for developing adaptive personalized interventions in SDLEs. SRTs provide answers to questions regarding treatment sequencing, timing, and personalization throughout a course. Answers to these questions will both improve outcomes for learners and deepen understanding of learning science in scalable environments.

In this work, we provided a few examples of different SRTs. There are more variations of SRTs that may be useful for SDLEs. In PERCS, all users have the same randomization for all three weeks. However, if a learner is responding well to treatment, it may not make sense to re-randomize them and change their current treatment, as is done in Sequential Multiple Assignment Randomized Trials (SMARTs) \cite{Murphy_2005, Pelham_2016}. Since many online courses are accessible to learners at any time, randomization timing could be based on when each user enters the course (randomized one, two, and three weeks from the day the user enrolls in a course). If treatment delivery timing is a research question, the delivery time can also be (re)randomized (e.g., each week of the course, re-randomize users to receive one weekly email or seven daily email reminders). Also, since many SDLEs provide learners with all course content at the beginning of the course, a trial aiming to re-randomize course content could use trigger-based re-randomizations--where users' future content is only changed after they have watched a certain video or completed a particular assignment--to prevent early exposure to future treatment.

Though useful, SRTs have limitations. SRTs are non-adaptive trial designs. If an experimenter wishes to both learn efficacious treatments and provide benefit to learners in the current trial, adaptive designs such as \cite{Williams_2018} would be more appropriate. Combining adaptive designs with SRTs may also be useful for both developing high-quality adaptive interventions and improving benefits for current learners \cite{Cheung_2015}.

PERCS also has clear limitations. For one, since some users (10\%) retook the online course multiple times, those users were repeated in multiple iterations of the trial. Secondly, it's important to note that PERCS was an exploratory trial not a confirmatory trial. The goal of PERCS was to explore several sequences of treatments and evaluate their efficacy for different learners. The next step for PERCS is to narrow down best treatment options based on current data (which may be different for Indian and US learners). Since the current evidence is not very strong, we would then run a second SRT with fewer treatments, acquiring more data on those treatments of interest. Once we have significant evidence to indicate which sequence of emails is optimal, we can then compare this learned optimal sequence of emails to a control in an A/B test confirmatory trial. Treatment A would be the hypothesized optimal adaptive sequence of emails and treatment B would be no email.

Additional analyses of the PERCS data would also be interesting. Using different outcomes other than course activity (such as course completion or  assignment performance) could help further understanding of intervention efficacy. Also,  email types had varying word lengths. Assessing how word length changes email efficacy could further elucidate the treatment effects.
Lastly, using additional learner demographic information such as age, gender, or previous education as potential moderators would be useful. Although we currently cannot collect that information through the course, other studies have demonstrated how these characteristics can be inferred from available data \cite{Brooks_2018}.

In the comparison of SRTs to A/B tests, we limited the comparison to one example design. There are many other possible comparators. For example, we could have compared PERCS to a trial which performs a single-randomization in week 2, instead of week 1. This new design would then allow one to discover mid-course moderators (measured prior to week 2). The new design, however, would not be able to assess treatment effects in week 1.  An important characteristic of SRTs is that all of the questions mentioned in Section~\ref{seq_time_pers} can be answered from data collected in one trial.

\section{Acknowledgements}
This work was supported in part under the Holistic Modeling of Education (HOME) project funded by the Michigan Institute for Data Science (MIDAS).
We would also like to acknowledge Dr. Joseph Jay Williams and Dr. Daniel Almirall for valuable feedback on this work.

\bibliographystyle{abbrv}
\bibliography{main.bib}

\end{document}